\documentclass[journal, twocolumn]{IEEEtran}
\ifCLASSINFOpdf
\else
\fi
\usepackage{hhline}
 \usepackage{amssymb,amsmath,epsfig,cite,url,multicol,multirow}
\usepackage{hyperref}
\usepackage{graphicx}
\usepackage{float}
\newcommand{\ra}[1]{\renewcommand{\arraystretch}{#1}}
\usepackage{tipa}
\begin{document}
\title{Consonant-Vowel Transition Models Based on Deep Learning for Objective Evaluation of Articulation}

\author{Vikram C. Mathad, Julie M. Liss,  Kathy Chapman,  Nancy Scherer, and Visar Berisha   
        
\thanks{Vikram C. Mathad is with zapr media labs, Bangalore, India, 560016, Email: cmvikramshiva@gmail.com. Julie M. Liss is with the Department of Speech \& Hearing Science, Arizona State University, Tempe, AZ-85281, Email: julie.liss@asu.edu.  Kathy Chapman is with the Department of Communication Sciences and Disorders, University of Utah, Salt Lake City, UT-84112, Email: kathy.chapman@health.utah.edu.
Nancy Scherer is with the Department of Speech \& Hearing Science, Arizona State University, Tempe, AZ-85281, Email: nancy.scherer@asu.edu.  Visar Berisha is with the College of Health Solutions, and School of Electrical, Computer, \& Energy Engineering, Arizona State University, Tempe, AZ-85281, Email: visar@asu.edu. This work is funded in part by NIH-NIDCR DE026252, NIH-NIDCD R01DC006859, and NIH R21DE023519. }}
%NIDCR (R21 DC009654), to the first and

%last authors and Kathy Chapman R21 DE023519
\maketitle
\begin{abstract}
Spectro-temporal dynamics of consonant-vowel (CV) transition regions are considered to provide robust cues related to articulation. In this work, we propose an objective measure of precise articulation, dubbed the objective articulation measure (OAM), by analyzing the CV transitions segmented around vowel onsets. The OAM is derived based on the posteriors of a convolutional neural network pre-trained to classify between different consonants using CV regions as input. We demonstrate the  OAM is correlated with perceptual measures in a variety of contexts including (a) adult dysarthric speech, (b) the speech of children with cleft lip/palate, and (c) a database of accented English speech from native Mandarin and  Spanish speakers.

\end{abstract}

\section{Introduction}
Consonant-vowel (CV) transitions refer to the abrupt movement in articulatory gestures as a speaker moves from consonants to vowels~\cite{stevens1981evidence}. While producing CV transitions the articulators make dynamic movements which result in rapid changes in the amplitude and spectrum of the acoustic signal. Perceptual and algorithmic experiments conducted by Stevens concluded that the information in the CV boundary regions is robust and provides invariant cues related to articulation~\cite{stevens1981evidence, stevens2002toward}. For example, several papers demonstrate that CV transitions are efficient cues for place of articulation recognition in both adult and child speakers~\cite{parnell1978influence,  ohde2006perception, hedrick1993effect}. 
The challenging nature of these articulatory movements amplifies the acoustic manifestation of structural or functional impairments associated with the speech production mechanism. To that end, acoustic measures extracted from CV transitions have been analyzed as correlates of articulatory precision and intelligibility in motor speech disorders (e.g., apraxia and dysarthria), structural anomalies (e.g., cleft palate (CP), syndrome-related disorders (e.g., Down syndrome and galactosemia), and sensory-based disorders (e.g., hearing impairment)~\cite{gibbon2007research,henningsson2008universal}.

\begin{figure*}[tbh]
	\vspace{-0.2cm}
	\centering
	
	\includegraphics[height=45mm,width=\linewidth]{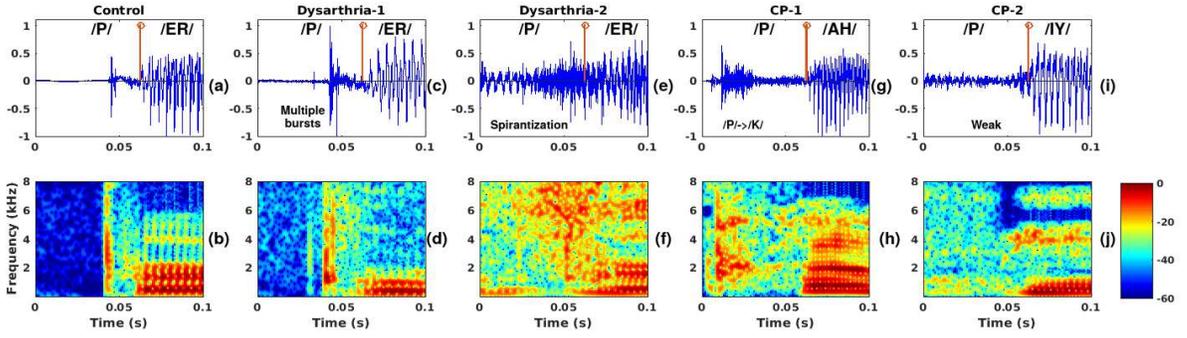}

	\caption{\label{CV1}  Different articulation errors produced for the target /p/. (a), (c), (e), (g), and (i) represent the waveforms of CV units containing target consonant \textipa{/p/} produced by control, dysarthric case-1, dysarthric case-2, CP case-1 and CP case-2, and the corresponding spectrograms are shown in (b), (d), (f), (h), and (j), respectively. For target \textipa{/p/}, the dysarthric case-1 produced multiple bursts and  case-2 produced spirantization error. CP case-1 substituted \textipa{/k/} for \textipa{/p/} and case-2 produced weak stop.  }

\end{figure*}

One of the challenges associated with assessment of speech acoustics during CV transition regions is signal variability. Different error types produce different acoustic patterns: devoicing and substitution errors in hearing impairment~\cite{osberger1982speech}; compensatory articulation errors (glottal and pharyngeal substitutions), and nasal substitutions in CP~\cite{henningsson2008universal}; nasalized consonants with longer duration in dysarthria~\cite{kim2010frequency, darley1969differential}; and phoneme substitution errors during second-language learning~\cite{tu2018investigating, witt2000phone} all differentially impact speech acoustics during CV transition regions. We demonstrate this with an example in Fig.~\ref{CV1}. In this figure we analyze the waveform and spectrogram of various types of articulation errors produced for the target stop consonant \textipa{/p/}. The waveform and spectrogram (Figs.~\ref{CV1}(a) and (b)) of the control speaker show a clear closure-burst transition for \textipa{/p/}. The dysarthric speakers show the presence of multiple bursts (Figs.~\ref{CV1}(c) and (d)) and spirantization (Figs.~\ref{CV1}(e) and (f)). \footnote{Spirantization refers to noise caused by incomplete articulatory closure during stop production.} The substitution error (\textipa{/p/}$\rightarrow$\textipa{/k/}) produced by the CP speaker shows the presence of a high-energy burst and increased VOT in Figs.~\ref{CV1}(g) and (h). The presence of noise in the consonant region (Figs.~\ref{CV1}(i) and (j)) indicates the presence of nasal air emission. The examples in Fig.~\ref{CV1} reveal that the acoustic characteristics of articulation errors during CV transitions vary across speech disorders.

\begin{figure*}[tbh]

	\centering
	
	\includegraphics[height=60mm,width=\linewidth]{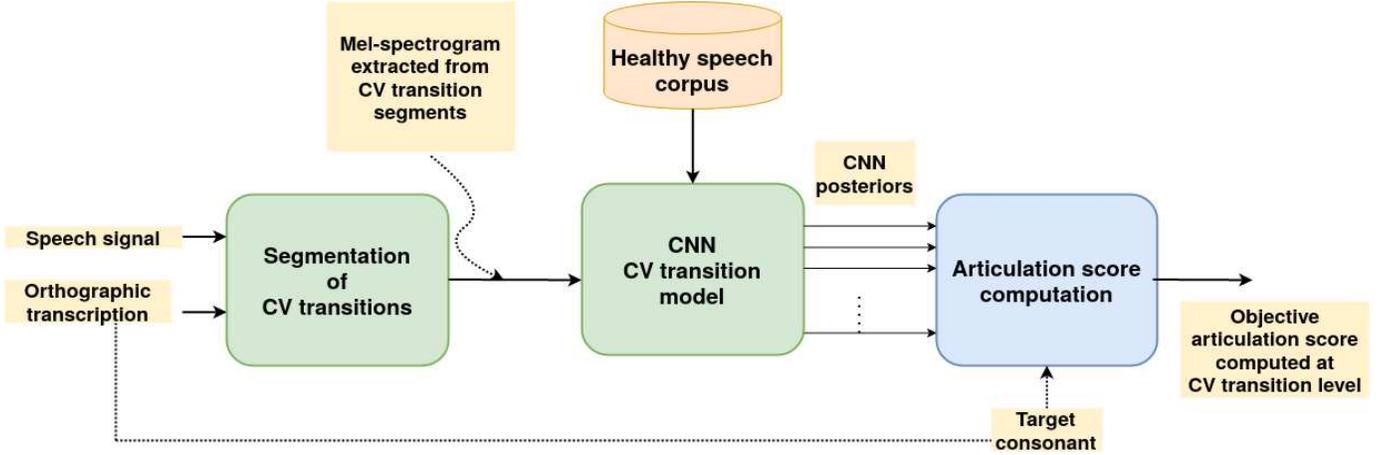}

	\caption{\label{bd} An overview of the system for computation of the objective articulation measure. }
	
\end{figure*}

This variability challenges the development of disease and speaker-invariant algorithms that assess CV transitions as objective proxies for precise articulation and intelligibility. The present work proposes an objective system for evaluation of CV transitions using an acoustic model pre-trained on a large normative speech sample. Consistent with existing works that tie CV transition errors to reduced articulatory precision, we show that the CV features produced by our model correlate with perceptual measures that capture articulation across a variety of speech disorders and accented speech.

\subsection{Related work}

%The perceptual and algorithmic experiments conducted by Steven~\cite{stevens1981evidence, stevens2002toward} concluded that the information in the CV boundary regions is robust and provides invariant cues for the perception of nasal and stop consonants. The perceptual experiments related to stop consonants showed the segment containing burst plus vocalic transition is the most efficient cue for place of articulation recognition in both adult and child speakers~\cite{parnell1978influence}. Nasal-to-vowel transition segments showed higher accuracy in the perception of place of articulation in both adult and child speaker groups than the nasal murmur region alone~\cite{ohde2006perception} Relative change between the frication noise and vowel onset amplitude was a robust cue for the perception of place of articulation in fricatives~\cite{hedrick1993effect}. 
The importance of CV transition regions, as highlighted by past experiments, has spurred the development of acoustic features for characterizing relevant CV features.  Traditionally, these features include a set of primitive acoustic measures including spectral peak, burst energy, spectral energy, spectral tilt, voice onset time (VOT), and formant transitions trajectories~\cite{kent1992acoustic, philips1984acoustic, kewley1983perception,sharf1972identification, sussman1993cross}. In many cases, these acoustic measures have been included as features in machine learning (ML) systems that objectively assess articulatory precision. These systems involve the supervised training of classifiers (e.g. support vector machine (SVM), Gaussian mixture model (GMM)) that map the features to clinical articulation scores. Most of these ML models~\cite{vikram2019detection, kalita2018intelligibility} are trained on a set of acoustic features extracted from a particular disorder. However, as we see in Fig.~\ref{CV1}, the nature of CV transition errors varies across speech disorders, consonant categories, and age groups. For example, VOT and formant transitions are used to assess oral-laryngeal coordination in the case of motor speech disorders~\cite{kent2003toward}, whereas low-frequency energy is used in the evaluation of nasalized consonants in CP~\cite{vikram2019detection,  philips1984acoustic}. Furthermore, the validity of several of these features critically depends on repeatable  estimation of formant frequencies and amplitudes, a notoriously difficult problem~\cite{stegmann2020repeatability}. These challenges make it difficult to develop models optimized for one disorder that generalize to other disorders.

As an alternative to traditional features based on signal processing, joint spectro-temporal features have been proposed to model CV transition regions in machine learning models. The two-dimensional discrete cosine transform (2D-DCT) is one of the most commonly used approaches to modeling the spectro-temporal dynamics of CV transition regions~\cite{karjigi2012classification, nossair1991dynamic, mathad2019vowel, kalita2018intelligibility}. Supervised learning methods using 2D-DCT features as input have been used for classification of place of articulation in stop consonants~\cite{nossair1991dynamic}, to evaluate the goodness of \textipa{/t/} and \textipa{/k/} productions in children with speech sound disorders~\cite{strombergsson2015acoustic}, to detect stop consonant production errors~\cite{mathad2019vowel}, and to model  perceptual intelligibility ratings in children with CP~\cite{kalita2018intelligibility}. Existing models based on supervised machine learning using the spectrum as input are trained for a very narrow context and are unlikely to generalize other contexts. Developing models that generalize to multiple contexts using this approach will require large and diverse labeled  clinical corpora for training~\cite{berisha2021digital}, a challenge too steep to overcome in many clinical applications.

\subsection{The present work}

Motivated by the significance of CV transitions and the limited availability of clinical speech data, in this work we propose a system for automated assessment of CV transition regions and demonstrate its validity by correlating the extracted features against perceptual ratings that capture articulatory precision across several clinical and non-clinical application domains. In contrast to clinical speech corpora, there are several publicly-available general population speech corpora. In our algorithm, we make use of the LibriSpeech corpus and train a discriminative CV feature learning model that learns to classify different consonants using CV regions as input. We use this pre-trained model as a CV feature extractor on input clinical speech samples. We further aggregate these features at the level of an utterance or speaker to generate an objective articulation measure (OAM). We demonstrate that the OAM correlates strongly ``out-of-the-box" with perceptual measures of articulation in several clinical speech applications and accentedness in speech from two different languages. That is, even without any supervised training using labels or speech samples from an application of interest, these features encode acoustic characteristics that correlate with human perception of properly produced speech. The performance of the OAM further improves with supervision if perceptual ratings are available for a corpus.

%%%%%%%%%%%%%%%%%%%%%%%%%%%%%%%%%%%%%%%%%%%

%
%

   \section{Databases}
   We use adult speech samples from the Librispeech corpus to build our CNN-based feature extractor and the OAM. We evaluate the OAM on three different corpora, two clinical corpora, and an accented speech database. Below we describe the details of the speech samples and perceptual ratings in each corpus.

   \subsection{Librispeech}
   The Librispeech database~\cite{panayotov2015librispeech} contains the utterances of adult American English speakers. In this work, a subset of the Librispeech database, {\it train-clean-100} and {\it test-clean}  are used in the development of the CNN. The training set contains the recordings of 
   251 adult speakers (125 male and 126 female).  A separate test set, comprised of 5.4 hours of speech, is used to evaluate the performance of the
   CNN classifier. 
   %The speech recordings and the corresponding  orthographic  transcriptions are force-aligned at the phoneme level using the Montreal force-aligner~\cite{mcauliffe2017montreal}.
   
    \subsection{Dysarthria}
   The Dysarthric speech database was collected at the Motor speech laboratory at Arizona State University ~\cite{saxon2019objective}.  The dysarthric speech database consists of speech recordings from 80 adult speakers (43 male and 37 female) of varying levels of articulation impairment: 38 participants with Parkinson's disease (PD), 6 participants with Huntington's disease (HD), 16 participants with cerebellar
   Ataxia (A), and 15 participants with amyotrophic lateral sclerosis
   (ALS), and 5 healthy controls. From each subject, a set of 5 sentences were recorded. The recordings were given to 14 speech-language pathologists (SLPs) to rate speaker-level articulatory precision on a scale of 1 to 7 (1-precise articulation, 7-severely imprecise articulation).   The average inter-rater correlation was found to be 0.831, $p<0.001$. For this study, we take the  average articulatory precision ratings of the 14 SLPs as the speaker-level articulatory precision score.  
   
   \subsection{Cleft speech database}
   The Utah Americelft cleft speech database is comprised of data from 60 children between the ages of 5-7 years old, each repeating 24 sentences from the Americleft speech evaluation protocol, recorded at the University of Utah~\cite{chapman2016americleft}. Speakers with CP produce obligatory (weak and nasalized consonants) and compensatory (glottal, pharyngeal, palatalized, and nasal fricative sounds) errors ~\cite{henningsson2008universal}. These errors lead to articulatory imprecision and degrade the acceptability of the produced speech. Speaker-level speech acceptability was rated on a  4-point scale (0-speech is acceptable, 1-speech is mildly unacceptable, 2-speech is moderately unacceptable, 3-speech is very unacceptable)  by 4 Americleft-trained SLPs, and the average correlation among the 4 raters was 0.776, $p<0.001$. For this study, we take the  average acceptability of the 4 SLPs as the speaker-level speech acceptability score.

     \subsection{Accented speech corpus}
   The accentedness database used in~\cite{tu2018investigating} was used as an additional test corpus. The accentedness database is a subset of the GMU speech accent archive~\cite{weinberger2015speech}. In this study, native (L1-language) Mandarin, and Spanish  speakers speaking American English (L2) are considered. The speaker set is the same as the subset used in the experiments of~\cite{tu2018investigating}, where for each language, 30 speakers, balanced on age and gender, were considered. Thirteen native American speakers were recruited as annotators to rate the accentedness on a 4-point rating scale: 1 = no accent/negligible accent, 2 = mild accent, 3 = strong accent, and 4 = very strong  accent. The average inter-rater correlation coefficient was found to be 0.73, $p<0.001$. The average of the ratings among the 13 annotators was computed and considered as the final accentedness score for all analyses in this paper.

\section{OAM computation}
  In Fig.~\ref{bd} we show a block diagram of the framework to compute the OAM. The OAM computation  involves (a) the segmentation of CV transition regions around the vowel onset, (b) training and validation of a CNN-based consonant classifier using CV transitions as an input, and  (c) articulation score computation. The different stages involved in the proposed system are described  in detail in the sections that follow.    

\begin{table}[b]
\centering
\caption{\label{CNN-details}Details of CNN architecture}
 \resizebox{\linewidth}{!}{
\begin{tabular}{cccc}
\\ \hline\hline
Sl. No. & Layer   & Activation & Dimension   \\ \hline
1       & Input   & -           & 40x32x1 \\
2       & Conv\_1 & ReLU       & 9x5x64  \\
3       & Pool\_1 &   -         & 2x2     \\
4       & Conv\_2 & ReLU       & 5x3x64  \\
5       & Pool\_2 &  -          & 2x2    \\
6       & FC\_1   & ReLU       & 1024    \\
7       & FC\_2   & ReLU       & 1024    \\
8       & FC\_3   & ReLU       & 1024    \\
9       & Output  & Softmax    & 20    \\
\hline\hline 
\end{tabular}
}
\end{table}

\subsection{CV Feature Extractor}
We use a pre-trained CNN classifier as a CV feature extractor. The CNN consonant classifier is trained to classify between different consonants spoken by healthy individuals using CV transition regions as input. The CNN is a context-independent consonant classifier that aims to classify among 20 consonants, i.e.,  plosives: \textipa{/b/}, \textipa{/d/}, \textipa{/g/}, \textipa{/p/}, \textipa{/t/}, \textipa{/k/},   fricatives: \textipa{/z/}, \textipa{/v/}, \textipa{/s/}, \textipa{/S/}, \textipa{/f/}, \textipa{/h/}, \textipa{/T/}, \textipa{/D/}, affricates: \textipa{/\textteshlig/}, \textipa{/\textdyoghlig/}, nasals: \textipa{/n/}, \textipa{/m/}, \textipa{/ng/},  and liquids: \textipa{/l/}, \textipa{/\*r/}. The hundred hours of healthy speech samples from the Librispeech database are used to train the CNN. The speech samples and  the corresponding  orthographic transcriptions are first passed through the Montreal forced alignment tool to obtain the phoneme boundaries~\cite{mcauliffe2017montreal}. From the forced-alignment, we determine the location of the vowel onsets. Around each vowel onset, we take a 160 ms window (80 ms to the right and 80 ms to the left of the vowel onset) to segment the CV transition region. We empirically justify the 160 ms window size in Section~\ref{sec-window}.  The  CV transition segment is analyzed using a Hamming window of 20 ms with a shift of 5 ms. From each frame, a 40-dimensional log Mel-filter bank is used to extract features, where the 40 filters are separated equally in Mel scale from 100 Hz to 7800 Hz. Finally, the segmented CV transition is represented by a 2-dimensional representation of dimension 40x32.

  The  mel-spectrogram  of CV transition regions is used as input to train the CNN consonant classifier.   The CNN architecture is described in Table~\ref{CNN-details}.  The model has two convolutional layers with max pooling, where 64 filters of shape 9x5 with  1x1 stride  and 64 filters of shape 5x3 with  1x1 stride are used in the first and second layers respectively.  In both layers, we use a rectified linear unit (ReLU) activation and a Max-pooling operator of shape 2x2 with a 1x1 stride. The output of convolutional layers is flattened and passed to  3 fully connected layers. Each layer has 1024 hidden neurons with a ReLU activation function. The output layer is comprised of 20 nodes with softmax activation, with each output corresponding to one of the 20 consonants. The error between true and predicted output values is evaluated using a categorical cross-entropy loss ($L$) given by
\begin{equation}
L=-\sum_{i=0}^{B-1}\sum_{j=0}^{M-1}y_j^{(i)}log(\hat{y}_j^{(i)}),
\end{equation}
where $B$ refers to the batch size, $M$ is the number of classes ($M=20$), and $y_j$  and $\hat{y}_j$ correspond to the ground truth and predicted outputs. In this work, we chose a batch size of 8 sentences (in each batch, we use the CV transitions from 8 sentences), and the Adam optimizer is used to minimize the error between ground truth and predicted values. The network is trained for 10 epochs and the learning rate is set to 0.001. 

\subsection{\label{sec-window}Setting the window length and evaluating the CNN }
\begin{figure}[tbh]
	
	\centering
	
	\includegraphics[height=70mm,width=\linewidth]{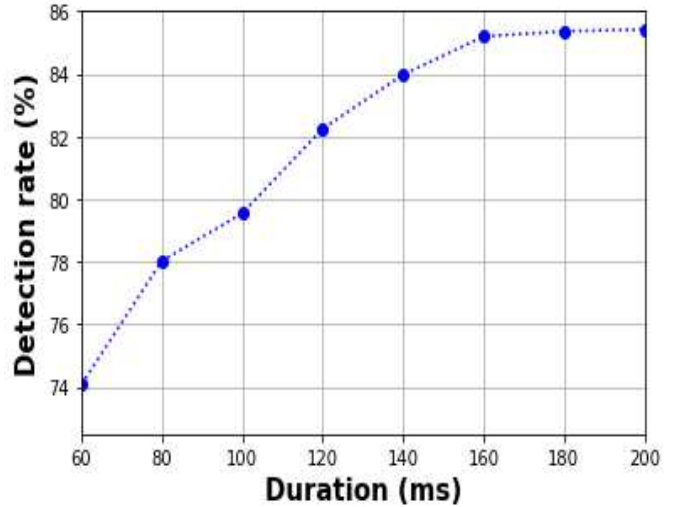}

	\caption{\label{CV-perf} Consonant detection rate for different CV transition duration. }

\end{figure}

We analyzed the dependency of the CNN classifier on CV transition duration. We vary the CV window around the vowel onset from 60 ms to 200 ms in steps of 20 ms and, for each case, we evaluate the consonant classification accuracy on the Librispeech test samples. Fig.~\ref{CV-perf} shows the CNN's performance as a function of the CV transition window size. From 60 to 160 ms, the performance increases with transition duration and then plateaus to an accuracy of $\sim$ 85\%. The improvement in performance is the result of improved model accuracy among consonants like affricates, which require a longer duration to produce and, therefore require longer windows to discriminate from others. For example, for a 60 ms transition duration, 47.212\% of instances of \textipa{/\textteshlig/} are falsely detected as \textipa{/S/}. Both are similar for place of articulation (palatal-alveolar) but differ in manner of production (fricative vs affricate). After increasing the window size to 160 ms, the error rate decreases to 13.283\%. The affricate textipa{/\textteshlig/} is a combination of silence and frication noise. If we consider a very small CV transition window, then only the frication noise is provided as input to the CNN. This is insufficient to discriminate between  \textipa{/\textteshlig/} and \textipa{/S/}, hence \textipa{/\textteshlig/} is confused with \textipa{/S/}.  

\begin{figure*}[tbh]

	\centering
	\includegraphics[height=55mm,width=\linewidth]{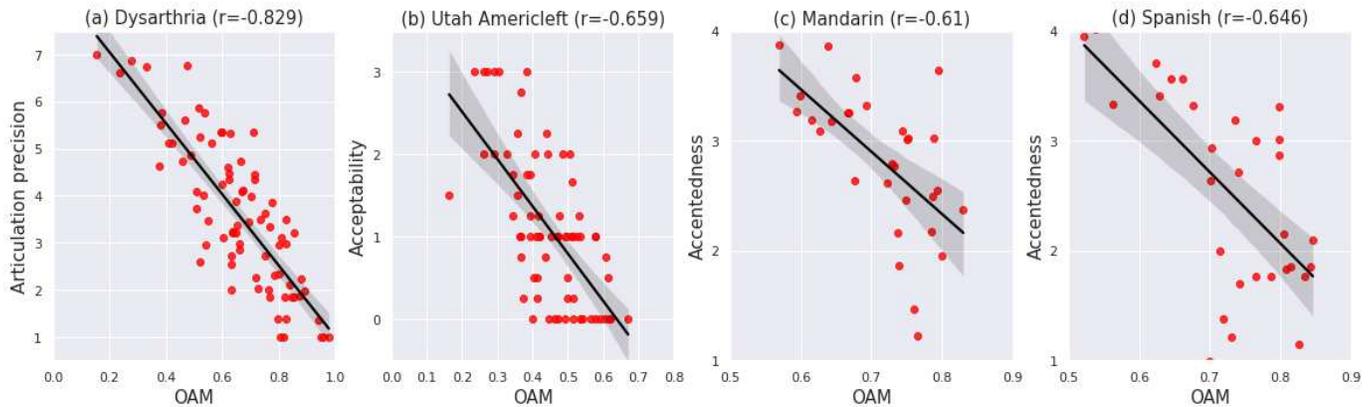}

	\caption{\label{Scatter_OAM} Scatter plots of speaker-level averaged OAM vs. perceptual ratings for  (a) dysarthria, (b) Utah Americleft, and (c) Mandarin and (d) Spanish databases.    }

\end{figure*}

\begin{figure*}[tbh]

	\centering
	\includegraphics[height=55mm,width=\linewidth]{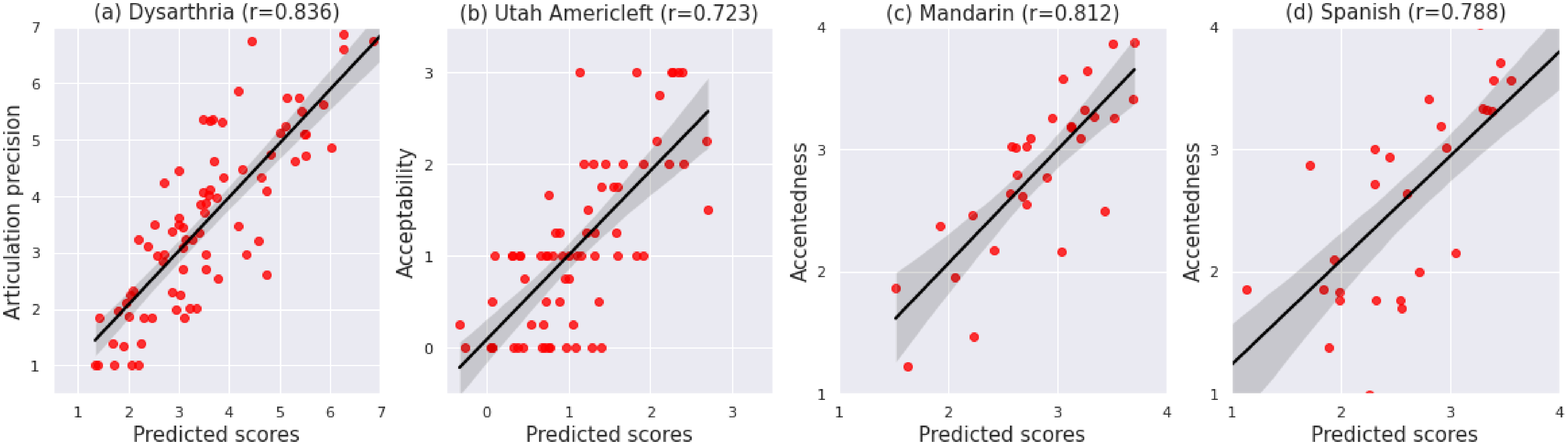}

	\caption{\label{Scatter_res} Scatter plots of predicted vs. perceptual ratings for  (a) dysarthria, (b) Utah Americelft, and (c) Mandarin and (d) Spanish databases.    }

\end{figure*}

\subsection{Articulation Score Computation}
The posteriors from the pre-trained CNN across the 20 classes are used as features for assessing consonant-level articulation scores in clinical speech and L2 speech. As shown in the block diagram Fig.~\ref{bd}, the input speech is first force-aligned to obtain the vowel onset locations. We extract a 160 ms CV segment from the input speech sample (as we did during training) and compute a mel-spectrogram. The mel-spectrogram is passed through the CNN to compute 20 posterior probability values from the pre-trained model. Let $c_i$ denote the segmented CV transition for the $i^{th}$ target consonant and let $P=\{p_0, p_1, p_2,...,p_{19}\}$ denote a vector of softmax layer outputs corresponding to the posterior probabilities of the 20 consonants. We define the OAM of consonant $c_i$, $OAM(c_i)$, as
\begin{equation}
OAM(c_i)=\frac{p_i}{max(P)}.
\label{eqn1}
\end{equation}  

For each speaker, the OAM is computed for all instances of target consonants. The vowel onset and target consonant information are obtained by using the Montreal forced-aligner~\cite{mcauliffe2017montreal}.

\vspace{0.2cm}
\noindent {\bf Consonant-level and speaker-level OAM scores:}
For each speaker,  OAM scores are computed for each consonant instance. Further, these scores  are averaged across all instances of a consonant for each speaker to obtain 20 consonant-level OAM scores for a speaker. These 20 consonant-level scores are averaged for each speaker to obtain a  speaker-level OAM. The OAM score achieves a maximum value of 1 when $p_i = max(P)$ for all consonants produced by a speaker.

\vspace{0.2cm}
\noindent {\bf Linear Models:}
For each speaker, we obtain 20 consonant-level OAM scores.  We develop domain-specific models of articulation using these 20 features to predict perceptual variables of interest in each domain: imprecise articulation in dysarthria, speech acceptability in CP, and accentedness in speech from L2-learners. The linear models are combined with a forward feature selection algorithm to learn a parsimonious model. The perceptual labels from each corpus are used as labels during training.

\section{Experimental Evaluation}

Herein we present the details of the experimental evaluation and results. First, we evaluate the relationship between the OAM and (a) articulation precision in dysarthria, (b) acceptability in children with CP, (c) accentedness in Mandarin, and (d) accentedness in Spanish speakers. Next, we analyze the effect of forced-alignment errors on the OAM and visualize the internal workings of the pre-trained DNN feature extractor via saliency maps. Finally, we compare the OAM with the goodness of pronunciation (GOP) algorithm, a commonly used algorithm for objective articulation assessment.

\subsection{Comparing the OAM scores to perceptual labels}

We demonstrate the validity of the OAM by comparing the acoustic-derived features with perceptual measures of articulatory precision in dysarthria, speech acceptability in cleft palate, and accentedness in Mandarin and Spanish speakers. We evaluate the correlation between acoustics and perceptual measures for two types of acoustic measures: (a) speaker-level  OAM scores and (b) linear predictors of perceptual labels using the OAM computed at the consonant level. The scatter plots of speaker-level OAM vs. perceptual ratings  for Dysarthria, Utah Americleft, Mandarin, and Spanish  are shown in  Fig.~\ref{Scatter_OAM}(a)-(d), respectively.   As expected, the higher the OAM, the closer to normal the articulatory precision ratings (1 on the scale). 

The linear model's performance is evaluated using the leave-one-speaker-out (LOSO) method. In LOSO validation, the consonant-level OAM scores of all speakers except one are used to train the linear model, and the held-out speaker is used during testing. A scatter plot between the predicted scores and perceptual ratings for Dysarthria, Utah Americleft, Mandarin, and Spanish are shown in Fig.~\ref{Scatter_res}(a)-(d), respectively.  

The results for the average OAM and the linear model are presented in Table~\ref{Res}. Across all four databases, the averaged OAM scores showed a significant correlation with the perceptual ratings, despite the fact that the training of the CNN does not require any perceptual labels. As expected, the linear model-based approach shows  improvement over the  average OAM.

\begin{table}[tbh]
	\centering
		\caption{\label{Res} Results of Speaker-level OAM    } 
		\ra{1.1}
		\resizebox{1\linewidth}{!}{
	\begin{tabular}{llll}
		\hline\hline
		Database   & Measure & \begin{tabular}[c]{@{}c@{}}Average\\ (r)\end{tabular} & \begin{tabular}[c]{@{}c@{}}Linear\\ (r)\end{tabular} \\ \hline
		Dysarthria & OAM     & -0.829  & 0.836  \\
		& $G_{C}$      & -0.795  & 0.796  \\
		& $G_{C+V}$     & -0.818  & 0.848  \\ \hline
		CP         & OAM     & -0.659  & 0.723  \\
		& $G_{C}$      & -0.689  & 0.725  \\
		& $G_{C+V}$     & -0.641  & 0.805  \\ \hline
		Mandarin   & OAM     & -0.609  & 0.812  \\
		& $G_{C}$      & -0.526  & 0.583  \\
		& $G_{C+V}$     & -0.597  & 0.730  \\ \hline
		Spanish    & OAM     & -0.645  & 0.788  \\
		& $G_{C}$      & -0.747  & 0.717  \\
		& $G_{C+V}$     & -0.744  & 0.727  \\ \hline\hline
	\end{tabular}
}
\end{table}

\subsection{The impact of alignment errors on the OAM}
Computation of the OAM requires detection of vowel onsets, which are obtained automatically using forced-alignment. Forced alignment algorithms can be error prone, i.e., there are timing differences between manual and automatic alignments~\cite{mahr2021performance}. To evaluate the impact of forced alignment error on the OAM, we evaluate the OAM scores using both automatic and manual alignments. Manual alignments are obtained by hand-correction of the forced alignments resulting from the Montreal aligner. The phoneme-boundaries of Praat-textgrid files are corrected by listening to the audio samples individually using Praat.  The distribution of alignment error for the different databases are shown in the second column of Table~\ref{AE}. This table also shows the correlation values for the predicted scores obtained for manual and automatic alignments. 

The results for automatic alignments are consistently greater than for the manual case. This is likely because the alignment error increases as the speech severity increases. Hence, the CV transition region segmented using incorrectly detected vowel onsets deviates from their target, and the alignment error contributes to the estimated OAM in addition to acoustic deviation induced by articulation error~\cite{mathad2021impact}.

\begin{table}[tbh]
	\centering
	\ra{1.4}
	\caption{\label{AE} Comparison of OAM for manual and automatic  alignments  } 
\begin{tabular}{cccc}
\hline\hline
Database   & \begin{tabular}[c]{@{}c@{}}Alignment error\\ (Mean+std) ms\end{tabular} & \begin{tabular}[c]{@{}c@{}}Manual\\ (r)\end{tabular} & \begin{tabular}[c]{@{}c@{}}Automatic\\ (r)\end{tabular} \\ \hline
Dysarthria & 22.502+28.501                                                           & 0.766                                                & 0.836                                                   \\
CP         & 32.891+41.429                                                           & 0.707                                                & 0.712                                                   \\
Mandarin   & 18.870+19.263                                                           & 0.824                                                & 0.868                                                   \\
Spanish    & 18.832+21.680                                                           & 0.605                                                & 0.802    \\
\hline\hline                                            
\end{tabular}
\end{table}

\subsection{Understanding the spectral characteristics the OAM captures}
\begin{figure*}[tbh]
	\vspace{-0.2cm}
	\centering
	
	\includegraphics[height=100mm,width=\linewidth]{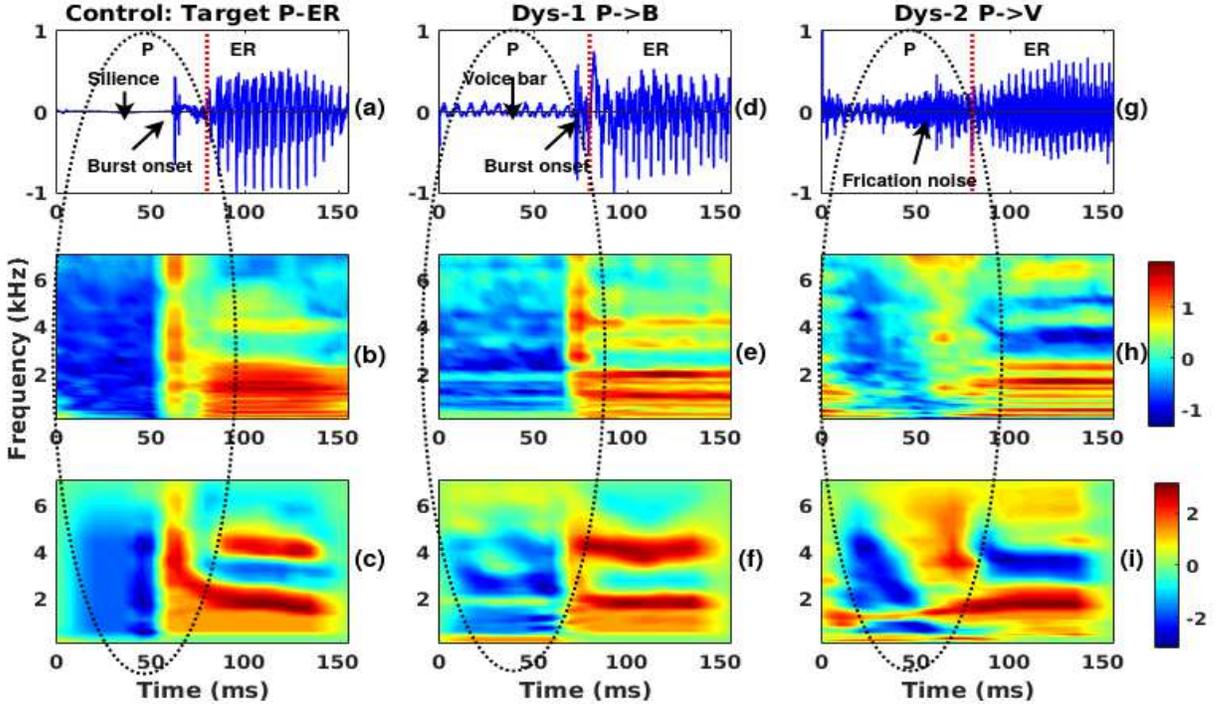}
	\vspace{-0.5cm}
	\caption{\label{sal} Saliency maps for the target CV transition . (a) waveform, (b) Mel-spectrogram, and (c) saliency map for control.  (d) waveform, (e) Mel-spectrogram, and (f) saliency map for dysarthric case-1 where \textipa{/p/} was detected as \textipa{/k/} by CNN.  (g) waveform, (h) Mel-spectrogram, and (i) saliency map for dysarthric case-2 where \textipa{/p/} was detected as \textipa{/v/} by CNN. The vowel onset is indicated by red dotted line each waveform plots ((a), (b), and (c)). The saliency maps high lights the burst onset in control and Dys-1 and formant transitions in control, Dys-1 and Dys-2 cases. The phonetic symbols in the plots are in the ARPABET encoded form. ARPABET to IPA mapping: P$\rightarrow$\textipa{/p/},  B$\rightarrow$\textipa{/b/}, V $\rightarrow$\textipa{/v/}, and ER$\rightarrow$\textipa{/\textschwa\textrhoticity/}   }
	
\end{figure*}

We used  guided backpropagation to obtain the saliency maps for visualizing the features relevant inside the pre-trained CNN feature detector~\cite{selvaraju2017grad}. Fig.~\ref{sal} shows the mel-spectrograms  and the corresponding saliency maps, obtained from the first convolutional layer, for healthy controls and dysarthric speakers for the target \textipa{/p/}. Fig.~\ref{sal}(a) shows the speech waveform produced by a healthy control for the CV transition \textipa{/p/}-\textipa{/\textschwa\textrhoticity/}; the corresponding mel-spectrogram and the saliency  maps are shown in Figs.~\ref{sal}(b) and (c), respectively. The mel spectrogram (Fig.~\ref{sal}(b)) indicates  a sharp increase in energy at the burst onset followed by formant transitions. The time-frequency components around the burst onset and the formant transitions around the vowel onset are highlighted in the saliency map (Fig.~\ref{sal}(c)) as relevant features. In  dysarthric case-1, the target \textipa{/p/} was detected as \textipa{/b/}. The place of articulation remains the same in \textipa{/p/} and \textipa{/b/}, however \textipa{/p/} contains a silence indicating the unvoiced manner and \textipa{/b/} contains a voice bar component corresponding to voicing. The saliency map indicates the presence of low-frequency components corresponding to the voice bar component of \textipa{/b/}.  In dysarthric case-2, \textipa{/p/} is substituted by \textipa{/v/}. Hence, the CNN recognizes the target \textipa{/p/} as \textipa{/v/}. The saliency map for \textipa{/p/}$\rightarrow$\textipa{/v/} does not indicate the presence of a burst component as in Figs.~\ref{sal}(c) and (f). Instead, the formant transitions from the frication to the vowel  (around 50 ms in Fig.~\ref{sal}(i)) are highlighted. This analysis provides evidence that the CNN learned the discriminate features across various consonants and hence, is able to capture the acoustic features associated with articulation errors during these regions. 

\subsection{Comparison between the OAM and the GOP}

The performance of the proposed approach is compared with the GOP algorithm~\cite{witt2000phone}. The GOP is a popular method for the assessment of phone-level articulatory precision in L2-learners. The phone-level likelihood scores computed from the automatic speech recognizer (ASR) are used for the computation of GOP, where the ASR is trained for  healthy controls. Since, the OAM is also computed using the acoustic model trained on healthy speakers, we compared the performance of the OAM with the GOP.

We used a pre-trained DNN-HMM acoustic model \footnote[2]{https://s3.amazonaws.com/kaldi-dnn-ali-gop-models/dnn\_model.tar.gz} for GOP computation. The DNN-HMM model uses $i$-vector speaker adaptation.  The model was trained using  Kaldi scripts~\cite{povey2011kaldi} and the CMU pronunciation dictionary~\cite{weide1998cmu}.  The GOP scores are computed using the scripts available at\footnote[3]{https://github.com/tbright17/kaldi-dnn-ali-gop}.

One important difference between the two algorithms is that the OAM is intended as a feature extractor for CV transitions only, whereas the GOP is computed for each phoneme. As a result, we perform the comparison  by separately considering the GOP of consonants and the GOP of vowels and consonants. $G_{C}$ refers to the GOP scores evaluated for consonants only and $G_{C+V}$ refers to the GOP computed for consonants and vowels. Similar to the OAM, the performance of the GOP is evaluated in two ways, i.e., averaged GOP scores across the different phonemes and a linear model-based prediction built for each corpus. The performance of the GOP and the OAM for different databases are shown in Table~\ref{Res}. As with the OAM, the GOP also shows improved performance for the linear model when compared to simple averaging.

The OAM shows better performance than  $G_{C}$ in dysarthria, Mandarin and Spanish databases; however, the performance is comparable in the case of the CP database. Both GOP and OAM are computed using the acoustic models trained on the same Librispeech corpus. The presence of an acoustic mismatch between the adult and CP databases (which is comprised of children) can affect the estimated GOP scores. However, the DNN acoustic model uses $i$-vector based speaker adaptation and that can reduce the impact of the adult-child acoustic mismatch. The OAM scores, on the other hand, are computed directly on the child speech  without any kind of speaker adaptation; still the performance is comparable to the DNN that uses $i$-vector speaker adaptation. This result provides further evidence for the invariance argument made in the literature regarding CV transitions. In the case of both adults and children, the perception of  place of articulation for nasals and stop consonants is more efficient for CV transitions than  consonant regions alone \cite{parnell1978influence}.

      \begin{figure*}[tbh]
	\vspace{-0.2cm}
	\centering
	\includegraphics[height=52mm,width=\linewidth]{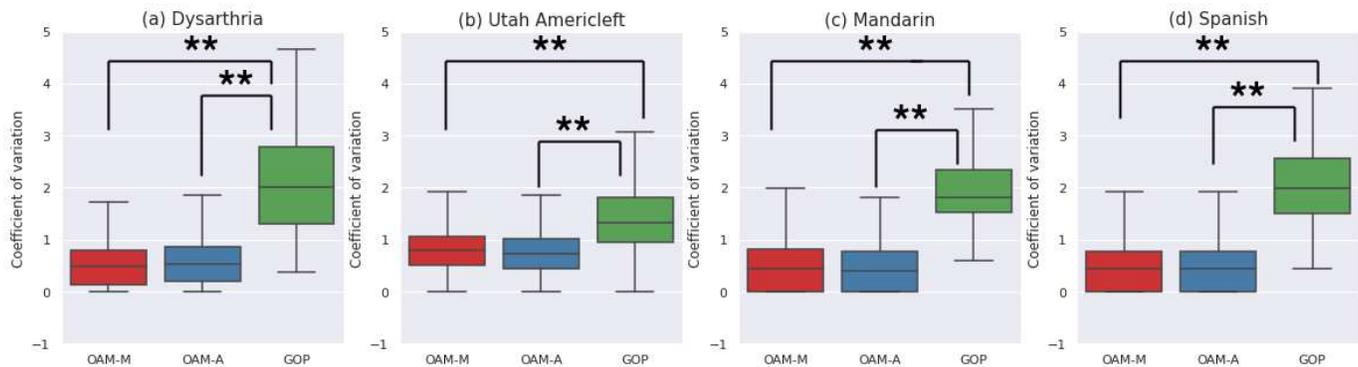}
	\vspace{-0.5cm}
	\caption{\label{CV} Boxplots of coefficient of variation for (a) dysarthria, (b) Utah Americleft, (c) Mandarin, and (d) Spanish databases ( $**p<0.001$)   }
	\vspace{-0.2cm}
\end{figure*}

\vspace{0.2cm}
\noindent {\bf Comparison of OAM and GOP variability:} Obtaining reliable GOP scores requires considerable averaging as the individual frame-level scores can be highly variable. This impacts the length of input speech required to obtain reliable estimates of the GOP. To evaluate the variability of the OAM relative to the GOP, we compare the coefficient of variation for the GOP and OAM across the four different databases considered.  The coefficient of variation  $\gamma_c$ for the phoneme $c$ is given by 
\begin{equation}
\gamma_c=\frac{\sigma_c}{| \mu_c|}
\end{equation}
where $\mu_c$ and $\sigma_c$ are the mean and standard deviation of the articulation scores estimated for various instances of the consonant $c$ per speaker. If the consonant category '$c$' occurs $N$ times for a speaker, then $\mu_c$ and $\sigma_c$ are estimated as,
\begin{equation}
\mu_c=\frac{1}{N}\sum_{i=0}^{N-1}A(c_i)
\end{equation}
\begin{equation}
\sigma_c=\sqrt{\frac{1}{N}\sum_{i=0}^{N-1}(A(c_i)-\mu_c)^2}
\end{equation}

Under the assumption that the articulation of a speaker is not variable, the ideal objective articulation score should yield low variance estimates across different instances of the same phoneme; hence, $\gamma_c $ is expected to be lower for more reliable objective measures. We compute $\gamma_c $ for  each consonant across all speakers in the database for the OAM and the GOP. Fig.~\ref{CV}(a)-(d) shows the distributions of $\gamma_c $ for the dysarthria, Utah Americelft, Mandarin, and Spanish databases, respectively. For each database, we plot the coefficient of variation for OAM-M, OAM-A, and the GOP. The coefficient of variation values are consistently higher for the GOP measure compared to the OAM.  A statistical evaluation using paired samples $t$-tests reveals that this difference is statistically significant. Across all databases, there is no statistically significant difference between the OAM-M and OAM-A ($p>0.001$). 

\vspace{0.2cm}
\noindent {\bf Computational complexity comparison: } During  computation of the GOP, the audio files and the corresponding orthographic transcriptions are passed through the pre-trained ASR model. Forced-alignment is used to obtain phoneme boundaries and phone-level likelihood scores are computed for the expected (target) and recognized phones to estimate the GOP score. To compute likelihood scores for each recognized phone, the algorithm needs to traverse across different triphone states of the acoustic model to estimate a  best possible path. Further, a language model is used to obtain more accurate sentence-level decoding using the recognized phoneme sequence. In addition, the DNN-based GOP computes 100-dimensional $i$-vectors and augments them with 40-dimensional MFCCs for speaker adaptation purposes. The computation of the OAM is much more straight-forward; it does not use any triphone decoding or language modeling. Instead, it only uses the posteriors of the expected and observed consonants. Hence, the computational complexity of the OAM is considerably smaller than the GOP. This results in a ~2X faster run time for the OAM computation relative to the GOP computation. We computed the execution time for a speech sample of length 5 sec. For the same file, the DNN-based GOP computation took 11 sec and the OAM computation took 4.811 sec (1.985 sec for Montreal forced aligner and 2.826 sec for the CNN).

\section{Summary and Conclusion}
In this work, we develop an algorithm for extracting features from CV transitions for the purpose of generating a general-purpose algorithm for the assessment of articulatory precision.  Motivated by previous work which indicates that CV transitions are largely context- and speaker-invariant, we develop a CNN consonant classifier to use as a feature extractor. The CNN developed in this work, captures both time and frequency variations of speech and  is trained using the mel-spectrogram computed from the CV transitions of healthy adult speakers. Analysis of saliency maps  shows that the model has learned important consonant cues like formant transitions and bursts in the transition regions to discriminate between the correct and incorrect articulation. We use the posteriors of the CNN to develop an objective articulation model (OAM) for prediction of articulatory precision in adult dysarthric speakers, speech acceptability in children with CP, and accentedness in L2-learners. The results show that the model output is significantly correlated with different perceptual scales across all databases. Furthermore, these correlations are on the order of those obtained using the GOP algorithm. However, in contrast to the GOP, the OAM model is less variable and computationally more efficient. One of the limitations of both the GOP and the OAM is that they require knowledge of the target transcript and force-alignment to segment individual phonemes (for the GOP) or CV transitions (for the OAM). Future work should focus on the development of objective articulation assessment models that are independent of transcription and forced-alignment operations.
   
 \bibliographystyle{IEEEtran}
 \bibliography{Stop_consonant_Errors}

\begin{thebibliography}{10}
\providecommand{\url}[1]{#1}
\csname url@samestyle\endcsname
\providecommand{\newblock}{\relax}
\providecommand{\bibinfo}[2]{#2}
\providecommand{\BIBentrySTDinterwordspacing}{\spaceskip=0pt\relax}
\providecommand{\BIBentryALTinterwordstretchfactor}{4}
\providecommand{\BIBentryALTinterwordspacing}{\spaceskip=\fontdimen2\font plus
\BIBentryALTinterwordstretchfactor\fontdimen3\font minus
  \fontdimen4\font\relax}
\providecommand{\BIBforeignlanguage}[2]{{%
\expandafter\ifx\csname l@#1\endcsname\relax
\typeout{** WARNING: IEEEtran.bst: No hyphenation pattern has been}%
\typeout{** loaded for the language `#1'. Using the pattern for}%
\typeout{** the default language instead.}%
\else
\language=\csname l@#1\endcsname
\fi
#2}}
\providecommand{\BIBdecl}{\relax}
\BIBdecl

\bibitem{stevens1981evidence}
K.~N. Stevens, ``Evidence for the role of acoustic boundaries in the perception
  of speech sounds,'' \emph{The Journal of the Acoustical Society of America},
  vol.~69, no.~S1, pp. S116--S116, 1981.

\bibitem{stevens2002toward}
------, ``Toward a model for lexical access based on acoustic landmarks and
  distinctive features,'' \emph{The Journal of the Acoustical Society of
  America}, vol. 111, no.~4, pp. 1872--1891, 2002.

\bibitem{parnell1978influence}
M.~M. Parnell, J.~D. Amerman, and C.~W. LaRiviere, ``Influence of speaker age
  on perceptual cue distribution,'' \emph{Journal of Phonetics}, vol.~6, no.~4,
  pp. 275--282, 1978.

\bibitem{ohde2006perception}
R.~N. Ohde, K.~L. Haley, and C.~W. Barnes, ``Perception of the [m]-[n]
  distinction in consonant-vowel (cv) and vowel-consonant (vc) syllables
  produced by child and adult talkers,'' \emph{The Journal of the Acoustical
  Society of America}, vol. 119, no.~3, pp. 1697--1711, 2006.

\bibitem{hedrick1993effect}
M.~S. Hedrick and R.~N. Ohde, ``Effect of relative amplitude of frication on
  perception of place of articulation,'' \emph{The Journal of the Acoustical
  Society of America}, vol.~94, no.~4, pp. 2005--2026, 1993.

\bibitem{gibbon2007research}
F.~E. Gibbon, ``Research and practice in developmental phonological
  disorders,'' in \emph{Phonology in context}.\hskip 1em plus 0.5em minus
  0.4em\relax Springer, 2007, pp. 245--273.

\bibitem{henningsson2008universal}
G.~Henningsson, D.~P. Kuehn, D.~Sell, T.~Sweeney, J.~E. Trost-Cardamone, and
  T.~L. Whitehill, ``Universal parameters for reporting speech outcomes in
  individuals with cleft palate,'' \emph{The Cleft Palate-Craniofacial
  Journal}, vol.~45, no.~1, pp. 1--17, 2008.

\bibitem{osberger1982speech}
M.~J. Osberger and N.~S. McGarr, ``Speech production characteristics of the
  hearing impaired,'' in \emph{Speech and Language}.\hskip 1em plus 0.5em minus
  0.4em\relax Elsevier, 1982, vol.~8, pp. 221--283.

\bibitem{kim2010frequency}
H.~Kim, K.~Martin, M.~Hasegawa-Johnson, and A.~Perlman, ``Frequency of
  consonant articulation errors in dysarthric speech,'' \emph{Clinical
  linguistics \& phonetics}, vol.~24, no.~10, pp. 759--770, 2010.

\bibitem{darley1969differential}
F.~L. Darley, A.~E. Aronson, and J.~R. Brown, ``Differential diagnostic
  patterns of dysarthria,'' \emph{Journal of speech and hearing research},
  vol.~12, no.~2, pp. 246--269, 1969.

\bibitem{tu2018investigating}
M.~Tu, A.~Grabek, J.~Liss, and V.~Berisha, ``Investigating the role of l1 in
  automatic pronunciation evaluation of l2 speech,'' in \emph{Proceedings of
  the Annual Conference of the International Speech Communication Association,
  INTERSPEECH}, vol. 2018, 2018, pp. 1636--1640.

\bibitem{witt2000phone}
S.~M. Witt and S.~J. Young, ``Phone-level pronunciation scoring and assessment
  for interactive language learning,'' \emph{Speech communication}, vol.~30,
  no. 2-3, pp. 95--108, 2000.

\bibitem{kent1992acoustic}
R.~D. Kent, C.~Read, and R.~D. Kent, \emph{The acoustic analysis of
  speech}.\hskip 1em plus 0.5em minus 0.4em\relax Singular Publishing Group San
  Diego, 1992, vol.~58.

\bibitem{philips1984acoustic}
B.~J. Philips and R.~D. Kent, ``Acoustic--phonetic descriptions of speech
  production in speakers with cleft palate and other velopharyngeal
  disorders,'' in \emph{Speech and Language}.\hskip 1em plus 0.5em minus
  0.4em\relax Elsevier, 1984, vol.~11, pp. 113--168.

\bibitem{kewley1983perception}
D.~Kewley-Port, D.~B. Pisoni, and M.~Studdert-Kennedy, ``Perception of static
  and dynamic acoustic cues to place of articulation in initial stop
  consonants,'' \emph{The Journal of the Acoustical Society of America},
  vol.~73, no.~5, pp. 1779--1793, 1983.

\bibitem{sharf1972identification}
D.~J. Sharf and T.~Hemeyer, ``Identification of place of consonant articulation
  from vowel formant transitions,'' \emph{The Journal of the Acoustical Society
  of America}, vol.~51, no.~2B, pp. 652--658, 1972.

\bibitem{sussman1993cross}
H.~M. Sussman, K.~A. Hoemeke, and F.~S. Ahmed, ``A cross-linguistic
  investigation of locus equations as a phonetic descriptor for place of
  articulation,'' \emph{The Journal of the Acoustical Society of America},
  vol.~94, no.~3, pp. 1256--1268, 1993.

\bibitem{vikram2019detection}
C.~Vikram, N.~Adiga, and S.~M. Prasanna, ``Detection of nasalized voiced stops
  in cleft palate speech using epoch-synchronous features,'' \emph{IEEE/ACM
  Transactions on Audio, Speech, and Language Processing}, vol.~27, no.~7, pp.
  1189--1200, 2019.

\bibitem{kalita2018intelligibility}
S.~Kalita, S.~Mahadeva~Prasanna, and S.~Dandapat, ``Intelligibility assessment
  of cleft lip and palate speech using gaussian posteriograms based on joint
  spectro-temporal features,'' \emph{The Journal of the Acoustical Society of
  America}, vol. 144, no.~4, pp. 2413--2423, 2018.

\bibitem{kent2003toward}
R.~D. Kent and Y.-J. Kim, ``Toward an acoustic typology of motor speech
  disorders,'' \emph{Clinical linguistics \& phonetics}, vol.~17, no.~6, pp.
  427--445, 2003.

\bibitem{stegmann2020repeatability}
G.~M. Stegmann, S.~Hahn, J.~Liss, J.~Shefner, S.~B. Rutkove, K.~Kawabata,
  S.~Bhandari, K.~Shelton, C.~J. Duncan, and V.~Berisha, ``Repeatability of
  commonly used speech and language features for clinical applications,''
  \emph{Digital biomarkers}, vol.~4, no.~3, pp. 109--122, 2020.

\bibitem{karjigi2012classification}
V.~Karjigi and P.~Rao, ``Classification of place of articulation in unvoiced
  stops with spectro-temporal surface modeling,'' \emph{Speech Communication},
  vol.~54, no.~10, pp. 1104--1120, 2012.

\bibitem{nossair1991dynamic}
Z.~B. Nossair and S.~A. Zahorian, ``Dynamic spectral shape features as acoustic
  correlates for initial stop consonants,'' \emph{The Journal of the Acoustical
  Society of America}, vol.~89, no.~6, pp. 2978--2991, 1991.

\bibitem{mathad2019vowel}
V.~C. Mathad and S.~M. Prasanna, ``Vowel onset point based screening of
  misarticulated stops in cleft lip and palate speech,'' \emph{IEEE/ACM
  Transactions on Audio, Speech, and Language Processing}, vol.~28, pp.
  450--460, 2019.

\bibitem{strombergsson2015acoustic}
S.~Str{\"o}mbergsson, G.~Salvi, and D.~House, ``Acoustic and perceptual
  evaluation of category goodness of/t/and/k/in typical and misarticulated
  children's speech,'' \emph{The Journal of the Acoustical Society of America},
  vol. 137, no.~6, pp. 3422--3435, 2015.

\bibitem{berisha2021digital}
V.~Berisha, C.~Krantsevich, P.~R. Hahn, S.~Hahn, G.~Dasarathy, P.~Turaga, and
  J.~Liss, ``Digital medicine and the curse of dimensionality,'' \emph{NPJ
  digital medicine}, vol.~4, no.~1, pp. 1--8, 2021.

\bibitem{panayotov2015librispeech}
V.~Panayotov, G.~Chen, D.~Povey, and S.~Khudanpur, ``Librispeech: an asr corpus
  based on public domain audio books,'' in \emph{2015 IEEE international
  conference on acoustics, speech and signal processing (ICASSP)}.\hskip 1em
  plus 0.5em minus 0.4em\relax IEEE, 2015, pp. 5206--5210.

\bibitem{saxon2019objective}
M.~Saxon, J.~Liss, and V.~Berisha, ``Objective measures of plosive nasalization
  in hypernasal speech,'' in \emph{ICASSP 2019-2019 IEEE International
  Conference on Acoustics, Speech and Signal Processing (ICASSP)}.\hskip 1em
  plus 0.5em minus 0.4em\relax IEEE, 2019, pp. 6520--6524.

\bibitem{chapman2016americleft}
K.~L. Chapman, A.~Baylis, J.~Trost-Cardamone, K.~N. Cordero, A.~Dixon,
  C.~Dobbelsteyn, A.~Thurmes, K.~Wilson, A.~Harding-Bell, T.~Sweeney
  \emph{et~al.}, ``The americleft speech project: a training and reliability
  study,'' \emph{The Cleft Palate-Craniofacial Journal}, vol.~53, no.~1, pp.
  93--108, 2016.

\bibitem{weinberger2015speech}
S.~Weinberger, ``Speech accent archive. george mason university,''
  \emph{Online:< http://accent. gmu. edu}, 2015.

\bibitem{mcauliffe2017montreal}
M.~McAuliffe, M.~Socolof, S.~Mihuc, M.~Wagner, and M.~Sonderegger, ``Montreal
  forced aligner: Trainable text-speech alignment using kaldi.'' in
  \emph{Interspeech}, vol. 2017, 2017, pp. 498--502.

\bibitem{mahr2021performance}
T.~J. Mahr, V.~Berisha, K.~Kawabata, J.~Liss, and K.~C. Hustad, ``Performance
  of forced-alignment algorithms on children's speech,'' \emph{Journal of
  Speech, Language, and Hearing Research}, pp. 1--10, 2021.

\bibitem{mathad2021impact}
V.~C. Mathad, T.~J. Mahr, N.~Scherer, K.~Chapman, K.~C. Hustad, J.~Liss, and
  V.~Berisha, ``The impact of forced-alignment errors on automatic
  pronunciation evaluation,'' 2021.

\bibitem{selvaraju2017grad}
R.~R. Selvaraju, M.~Cogswell, A.~Das, R.~Vedantam, D.~Parikh, and D.~Batra,
  ``Grad-cam: Visual explanations from deep networks via gradient-based
  localization,'' in \emph{Proceedings of the IEEE international conference on
  computer vision}, 2017, pp. 618--626.

\bibitem{povey2011kaldi}
D.~Povey, A.~Ghoshal, G.~Boulianne, L.~Burget, O.~Glembek, N.~Goel,
  M.~Hannemann, P.~Motlicek, Y.~Qian, P.~Schwarz \emph{et~al.}, ``The kaldi
  speech recognition toolkit,'' in \emph{IEEE 2011 workshop on automatic speech
  recognition and understanding}, no. CONF.\hskip 1em plus 0.5em minus
  0.4em\relax IEEE Signal Processing Society, 2011.

\bibitem{weide1998cmu}
R.~Weide, ``The cmu pronunciation dictionary, release 0.6,'' 1998.

\end{thebibliography}

\end{document}